\documentclass[aps,pra,amssymb,amsfonts,amsmath,showpacs]{revtex4}

\begin{document}

\title{Quasirelativistic quasilocal finite wave-function collapse model}
\author{Philip Pearle}
\email{ppearle@hamilton.edu}
\affiliation{Department of Physics, Hamilton College, Clinton, NY  13323}
\date{\today}

\begin{abstract}
{A Markovian wave function collapse model  is presented where the collapse-inducing operator, constructed from quantum fields,  is  a manifestly covariant 
generalization of the mass density operator utilized in the nonrelativistic 
Continuous Spontaneous Localization (CSL) wave function collapse model.  
However, the model is not Lorentz invariant because two such operators 
do not commute at spacelike separation, i.e., the time-ordering operation in 
one Lorentz frame, the ``preferred" frame, is not the time-ordering operation in another frame.
However, the characteristic spacelike distance over which the commutator decays is the particle's
Compton wavelength so, since the commutator rapidly gets quite small, the 
model is ``almost" relativistic. This ``QRCSL"  model is completely finite:  unlike 
previous, relativistic, models, it has no (infinite) energy production from the vacuum state. 
 QRCSL calculations are given of 
the collapse rate for a single free particle in a superposition of spatially separated packets, and of the energy production rate 
for any number of  free particles: these reduce to 
the CSL rates if the particle's Compton wavelength is small 
compared to the model's distance parameter.   One motivation for QRCSL is the realization that previous  relativistic models entail excitation 
of nuclear states which exceeds that of experiment, whereas QRCSL does not: an example is given involving quadrupole 
excitation of the $^{74}$Ge nucleus.}
\end{abstract}


\maketitle

\section{Introduction}\label{Section I}

	The Continuous Spontaneous Localization (CSL) model\cite{PearleCSL,GPR} 
is, currently, the only modification of Schr\"odinger's equation which satisfactorily describes 
both standard quantum physics and the macroscopic world we see around us.  When a 
superposition of macroscopic spatially distinguishable states appears, the wave function dynamically 
undergoes rapid collapse toward one such state.  

	Since the world is locally Lorentz invariant, it is natural to try to make a relativistic collapse 
model.  I have been working at this for over a decade and, although the models constructed have 
certain interesting features, each has certain interesting flaws.  

The major problem with the first 
models\cite{PearleRel,GGP,PearleRel2}	is that, even to lowest order in the collapse rate parameter $\lambda$, there is 
an infinite energy production rate per volume from the vacuum.  
The reason is as follows. In nonrelativistic CSL, 
a randomly fluctuating classical field $w({\bf x}, t)$ interacts with the mass-density 
operator (coupling constant  $\lambda$).  Collapse narrows wavepackets, 
resulting in a small rate of energy increase of particles (the 
energy is supplied by $w({\bf x}, t)$\cite{Pearleenergy}) which is, at present, below experimental observation\cite{experiment}.  
However, in these relativistic models, each vacuum energy-momentum mode, which may be regarded as the ground state of a harmonic oscillator,  is likewise narrowed by the collapse interaction.  Thus each  
vacuum mode is no longer just the ground state but is a superposition of ground and excited states,   
 i.e., there is a small rate of creation of particles with that energy-momentum.  
Since there are an infinite number of modes, than infinite energy production rate per volume of particles from the vacuum. 

	In nonrelativistic CSL and in these first relativistic models, $w({\bf x}, t)$ is white gaussian noise, 
i.e., it contains all wavelengths and frequencies in equal amounts.  In standard 
quantum theory, interaction of an operator with classical noise of frequency $\omega$ 
results in excitation of the quantum system with energy change $\hbar\omega$.  That is also the case for 
collapse models.  

	I therefore considered models where the noise is non-white gaussian\cite{PearleNoise, BGNoise} (which entails 
a non-Markovian state vector  evolution) and showed that, indeed, suppression of 
the frequency $\omega$ in the {\it inverse} of the spectrum of $w({\bf x}, t)$ suppresses 
energy $\hbar\omega$ excitation.  This led to a relativisitic 
model with a tachyonic (momentum-energy relation ${\bf k}^{2}-k^{0}\thinspace^{2}=-\mu ^{2}$) inverse  noise spectrum\cite{PearleTachyon}, which 
does not excite the vacuum to lowest order in $\lambda$ since there is no  mode of the vacuum which has  tachyonic energy-momentum.  

	However, to order $\lambda^{2}$, 
once again the ugly spectre(um) of infinite energy production from the vacuum 
appears. The  culprit is that, in this order, there is an internal particle 
line in the Feynman diagram describing vacuum excitation.  The associated 
particle propagator, $({\bf k}^{2}-k^{0}\thinspace^{2}+M ^{2}-i\epsilon)^{-1}$, like the white noise spectrum, 
possesses all wavelengths and energies, which it uses to convert the tachyonic energy-momentum 
to vacuum production of a pair of particles. 

	If the particle propagator were on-shell,   
$\delta({\bf k}^{2}-k^{0}\thinspace^{2}+M ^{2})$, then the vacuum excitation disappears 
 to this and all orders.  This may be achieved by removing the time-ordering operation from 
the state-vector evolution operator, resulting in a finite relativistically invariant 
collapse model (RCSL), but at a cost.  First, the model is expressed in the interaction picture: with time-ordering one can reconstruct the Schr\"odinger picture (multiply the interaction picture state vector by 
$\exp-iHt$, where $H$ is the Hamiltonian) 
but, without time-ordering, this connection is severed.  Second, the model is nonlocal in an 
unusual sense since, with time ordering, each Feynman diagram describes a series of forward-in-time 
evolutions but, without time-ordering, these alternate with backward-in-time evolutions.  While the time-ordering evolution keeps an evolving particle rigorously within the light cone of its initial spatial state, the non-time-ordering evolution does allow the particle to go out of the light cone in high enough order, albeit with a small probability.  Third, and most decisively, as shown in Section V and Appendix A, to lowest order in $\lambda$ (where time-ordering plays no role), the tachyonic-based theory predicts too great an excitation of nuclear states: the ``spontaneous" quadrupole excitation from the ground state  ($0^{+}$) of $^{74}$Ge to the first excited state ($2^{+}$) at .596MeV greatly exceeds the experimentally observed  rate.  To my knowledge, this is the first situation where experiment has ruled out a collapse model.  

	This paper takes another tack.  I have been unable to to construct a collapse model 
simultaneously satisfying 1) relativistic invariance, 2) locality, 3) Markovian evolution, 4) no vacuum excitation, 5) consistency with experiment. Nonrelativistic CSL 
satisfies all  but 1), replacing it with galilean invariance.  Relativistic collapse models in refs\cite{PearleRel,GGP,PearleRel2} do not satisfy 4), 5) and RCSL does not satisfy 2), 3) and 5). (See references\cite{RN,Tumulka} for a similar assessment of recent interesting models by Rimini and Nicrosini and by Tumulka.)  This paper contains a model which satisfies 3), 4), 5), which satisfies 
1) and 2) approximately, and which reduces to  CSL in the nonrelativistic limit. 

\section{Nonrelativistic CSL}\label{Section II}

	Nonrelativistic CSL is characterized by two equations, the  state vector evolution equation and the 
probability rule. 

	The evolution equation in the ``collapse interaction picture" (where the operators are Heisenberg operators, and the state vector only changes with time due to collapse) is	
\begin{equation}\label{1}
	|\psi , t \rangle_{w}={\cal T} e^{-(4\lambda)^{-1}\int_{0}^{t}dt d{\bf x}[w({\bf x},t)-2\lambda A({\bf x},t)]^{2}}|\psi ,0 \rangle
\end{equation}
\noindent (${\cal T} $ is the time-ordering operator).  In Eq.(1), $A({\bf x},t)$ is an operator essentially proportional to the 
mass of particles in a spherical volume of radius $a$, and can be written in various ways:
\begin{subequations}
\begin{eqnarray}\label{2}
A({\bf x},0)&\equiv& \sum_{\alpha} \frac{M_{\alpha}}{M}(4\pi a^{2})^{3/4}e^{2^{-1}a^{2}\nabla^{2}}
\xi_{\alpha}^{\dagger}({\bf x})\xi_{\alpha}({\bf x})\label{2a},\\
&=& \sum_{\alpha} \frac{M_{\alpha}}{M}\frac{1}{(\pi a^{2})^{3/4}}\int d{\bf b}e^{-(2a^{2})^{-1}{\bf b^{2}}}
\xi_{\alpha}^{\dagger}({\bf x}+{\bf b})\xi_{\alpha}({\bf x}+{\bf b})\label{2b}\\
&=&\sum_{\alpha} \frac{M_{\alpha}}{M}\bigg(\frac{a^{2}}{4\pi^{3}}\bigg)^{3/4}\int d{\bf p}'d{\bf p}e^{-2^{-1}
a^{2}({\bf p}'-{\bf p})^{2}}
e^{-i({\bf p}'-{\bf p})\cdot {\bf x}}a_{\alpha}^{\dagger}({\bf p}')a_{\alpha}({\bf p}).\label{2c} 
\end{eqnarray}
 \end{subequations}
\noindent where $ A({\bf x},t)=\exp (iHt) A({\bf x},0)\exp (-iHt)$, with $H$ the complete Hamiltonian for the interacting particles.  In Eqs.(2), 
\begin{equation} \label{3}
 \xi_{\alpha}^{\dagger}({\bf x})\equiv (2\pi )^{-3/2}\int d{\bf p}e^{-i{\bf p}\cdot {\bf x}}a_{\alpha}^{\dagger}({\bf p})
\end{equation}
\noindent is the creation operator for a particle of  type $\alpha$ (electron, proton, neutron,...) at the position ${\bf x}$ 
(so $\xi_{\alpha}^{\dagger}({\bf x})\xi_{\alpha}({\bf x})$ is the number density operator for particles of  type $\alpha$),   $a_{\alpha}^{\dagger}({\bf p})$
is its momentum ${\bf p}$ creation operator,  $M_{\alpha}$ is its mass and  $M$ is the proton mass.
The values of the parameters $\lambda$ and 
$a$ which characterize the model are generally chosen to be those given in the seminal 
collapse model of Ghirardi, Rimini and Weber (GRW) \cite{GRW}, $\lambda=10^{-16}$sec$^{-1}$ and 
$a=10^{-5}$cm, but there is a range of parameter values allowed by experiment\cite{Collett}.  

	The probability rule gives the probability density that the classical noise field  w({\bf x},t) 
occurs in nature:
\begin{equation}\label{4}
{\cal P}(w,t)\equiv \thinspace_{w}\langle\psi,t|\psi,t\rangle_{w}.
\end{equation}
\noindent That is,  Eq.(1) does not preserve statevector norm, and Eq.(4) says that state vectors of large 
norm are most likely to occur. It follows from Eqs.(1), (4) that $\int Dw{\cal P}(w,t)=1$, where 
$Dw\equiv\prod_{{\bf x},t}dw({\bf x},t)(2\pi \lambda/d{\bf x}dt)^{-1/2}$ is the functional integration volume element (in doing the integrals, ${\bf x}$, $t$ are discretized, with
$w({\bf x},t)$ regarded as an independent variable for each (${\bf x}$, $t$).

	It is readily shown that Eqs.(1), (4) entail that a state vector, describing a macroscopic object in a superposition of places, rapidly evolves toward one of the states in the superposition with probability equal to the squared magnitude of 
its coefficient in the superposition.  Essentially, a state which survives when all others have collapsed is  one for which the time average of $w({\bf x},t)$ equals the state's time average of $2\lambda A({\bf x},t)$ at each ${\bf x}$ (all other behaviors have vanishing probability).  
 
 	For calculations of physical effects, it is easiest to utilize the density matrix which describes the ensemble of evolutions:
\begin{subequations}
\begin{eqnarray}\label{5} 
	\rho (t)\equiv&&\int Dw \frac{|\psi, t\rangle_{w}\thinspace_{w}\langle \psi, t|}{\thinspace_{w}\langle \psi, t|
\psi, t\rangle_{w}}{\cal P}(w,t)\\
    =&&{\cal T} e^{-(\lambda/2)\int_{0}^{t}dtd{\bf x}[A_{L}({\bf x}, t)-A_{R}({\bf x}, t)]^{2}}\rho (0),
\end{eqnarray}
\end{subequations}
\noindent Eq.(5b) follows from putting Eqs.(1), (4) into (5a).  In Eq.(5b), when the exponential is expanded in a power series, $A_{L}$ ($A_{R}$) operates to the 
left (right) of $\rho (0)$, and ${\cal T}$ time-orders (time-reverse orders) the operators at the left (right).

\section{Quasirelativistic Model for Noninteracting Particles}\label{Section III}

	In what follows, for simplicity,  only one type of particle, a ``bosonic nucleon" of mass M shall be considered: 
the results are trivially extendable to fermions and bosons of many types. The relativistic generalization of the 
nonrelativistic creation operator  $\xi^{\dagger}({\bf x, t})$ for noninteracting particles (Eq.(3) where 
$\exp iEt$ is stuck into the integral, with $E={\bf p}^{2}/2M$) is the negative frequency field operator 
\begin{equation}\label{6}
\phi^{-}({\bf x, t})\equiv (2\pi )^{-3/2}\int d{\bf p}(M/E)^{1/2}e^{-ip\cdot x}a^{\dagger}({\bf p})
\end{equation}
\noindent ($p\cdot x\equiv {\bf p}\cdot{\bf x}-Et$, where $E=({\bf p}^{2}+M^{2})^{1/2}$) which, like 
the local field $\phi(x)=\phi^{-}(x)+\phi^{+}(x)$, transforms like a scalar under Lorentz transformations.  In 
Eq.(6) we choose the commutation relation $[a({\bf p}), a^{\dagger}({\bf p}')] =\delta ({\bf p}-{\bf p}')$ so  
$a({\bf p})\sqrt{E}$ transforms like a scalar.  It is clear that $\phi^{-}({\bf x, t})$ reduces to 
 $\xi^{\dagger}({\bf x, t})$ in the nonrelativistic ($c\rightarrow\infty$) limit (except for an additional factor $\exp iMt$, which cancels out in $\phi^{-}\phi^{+}$).  
 
 	In this quasirelativisitic CSL (QRCSL) model, the evolution equations are also (1) for the state vector,  (5)  for the 
density matrix, and (3) for the probability rule. But, for QRCSL, $A({\bf x, t})$ is defined using an approach from nonlocal relativistic quantum 
field theory\cite{Namsrai}, and can be written in various ways parallel to Eqs.(2):
\begin{subequations}
\begin{eqnarray}\label{7}
A(x)&\equiv&(4\pi a^{2})^{3/4}e^{2^{-1}a^{2}\Box }\phi^{-}(x)\phi^{+}(x)\\
&=&\frac{1}{2^{1/2}(\pi a^{2})^{5/4}}\int d{\bf b}db_{0}e^{-(2a^{2})^{-1}[{\bf b}^{2}+b_{0}^{2}]}
\phi^{-}({\bf x}+{\bf b}, t + ib_{0})\phi^{+}({\bf x}+{\bf b}, t + ib_{0})\label{7b}\\
&=&\bigg(\frac{a^{2}}{4\pi^{3}}\bigg)^{3/4}\int d{\bf p}'d{\bf p}\frac{M}{\sqrt{EE'}}e^{-2^{-1}
(p'-p)^{2}a^{2}}
e^{-i( p'-p)\cdot  x}a^{\dagger}({\bf p}')a({\bf p})\label{7c} 
\end{eqnarray}
 \end{subequations}
\noindent ($\Box \equiv\nabla^{2}-\partial_{t}^{2}$).

	It is apparent from Eqs.(7a,c) that $A(x)$ is a Lorentz scalar.  It is also worth noting that the exponent 
$(p'-p)^{2}$ in (7c) is spacelike (i.e., positive:  in the reference frame where ${\bf p}'=0$, $(p'-p)^{2}=2M(E-M)$).  
It is easy to see from Eqs.(7) that, in the $c\rightarrow\infty$ limit, A(x) reduces to the nonrelativistic  $A({\bf x},t)$ (Eqs.(2) with time behavior added).

	The equal time commutator, utilizing Eq.(7b), is		
\begin{eqnarray}\label{8}
	&&[A({\bf x},t),A({\bf x}',t)]=(2\pi^{5/2} a^{5})^{-1}\int d{\bf b}db_{0}d{\bf b}'db_{0}' 
e^{-(2a^{2})^{-1}[{\bf b}^{2}+b_{0}^{2}]}e^{-(2a^{2})^{-1}[{\bf b}'^{2}+b_{0}'^{2}]}\nonumber\\
&&\qquad[\phi^{-}({\bf x}+{\bf b}, t + ib_{0})\phi^{+}({\bf x}'+{\bf b}', t + ib_{0}')-
\phi^{-}({\bf x}'+{\bf b}', t + ib_{0}')\phi^{+}({\bf x}+{\bf b}, t + ib_{0})]\nonumber\\
&&\qquad\qquad[\phi^{+}({\bf x}+{\bf b}, t + ib_{0}), \phi^{-}({\bf x}'+{\bf b}', t + ib_{0}')]
\end{eqnarray}
\noindent with  
\begin{equation}\label{9}
[\phi^{+}(x), \phi^{-}(x')]=\frac{M}{(2\pi)^{3}}\int\frac{d{\bf p}}{E}e^{ip\cdot(x-x')}.
\end{equation}
\noindent Eq. (9),  with the arguments appropriate to (8), is
\begin{equation}\label{10}
[\phi^{+}({\bf x}+{\bf b}, t + ib_{0}), \phi^{-}({\bf x}'+{\bf b}', t + ib_{0}')]=
\frac{M^{2}K_{1}[M\sqrt{({\bf x}+{\bf b}-{\bf x}'-{\bf b}')^{2}+(b_{0}-b_{0}')^{2}}]}
{2\pi^{2}\sqrt{({\bf x}+{\bf b}-{\bf x}'-{\bf b}')^{2}+(b_{0}-b_{0}')^{2}}}.
\end{equation}
\noindent where $K_{1}$ is the Bessel Function.  Now, ${\bf b}, {\bf b}', b_{0}, b_{0}'$ in Eq.(8) have gaussian distributions 
with spread $a$.  For most of their volume of integration where the gaussians are large, the argument of $K_{1}$ 
in (10) is quite large, of order $Ma=10^{-5}$cm$/10^{-14}$cm$=10^{9}$.  Since for large argument, 
$K_{1}(z)\rightarrow(\pi/2z)^{1/2}\exp -z$, in Eq.(8) where the gaussians are large, the factor $K_{1}$ 
is, for the most part, quite small. It is in this sense that the commutator (8) ``almost" vanishes, making the 
time-ordering operation ``almost" frame-independent and the model ``quasi"-relativistic.  

	In calculations of the density matrix using Feynman diagrams, internal particle lines are 
represented by 	
\begin{subequations}
\begin{eqnarray}\label{11}
\langle 0|{\cal T}[\phi^{+}(x), \phi^{-}(x')]|0\rangle&=&\Theta (t-t')\frac{M}{(2\pi)^{3}}\int\frac{d{\bf p}}{E}e^{ip\cdot(x-x')}\\
&=&\frac{Mi}{(2\pi)^{4}}\int \frac{d^{4}p}{E}\frac{1}{p^{0}-E+i\epsilon}e^{ip\cdot(x-x')}\\
&=&\frac{2Mi}{(2\pi)^{4}}\int d^{4}p \frac{1+[(p^{0}-E)/2E]}{p^{02}-{\bf p}^{2}-m^{2}+i\epsilon}e^{ip\cdot(x-x')}.	
\end{eqnarray}
 \end{subequations}	
\noindent Eq.(11a) follows from Eq.(9): in (11a), $p=({\bf p},E)$ while in (11b,c), $p=({\bf p},p^{0})$.

	Eqs.(11) show the lack of Lorentz invariance.  However, with $({\bf x}-{\bf x}', t-t')$ replaced by 
$({\bf x}+{\bf b}-{\bf x}'-{\bf b}', t+ib_{0}-t'-ib_{0}')$ in (11a), the integrands of (11a,b,c) acquire the factor 
$\exp i[{\bf p}\cdot ({\bf b}-{\bf b}')+E(b_{0}-b_{0}')]$:  
for spacelike $(x-x')$, the propagator equals $\Theta (t-t')$ multiplied by 
Eq.(10) with argument $[M{({\bf x}+{\bf b}-{\bf x}'-{\bf b}')^{2}-(t+ib_{0}-t'-ib_{0}')^{2}}]^{1/2}$), 
and the previous discussion of how this ``almost" vanishes applies.  

	One might consider writing the propagator in each Feynman diagram, and thus the whole 
density matrix, as the sum of a relativistic piece plus a non-relativistic correction.  There 
are various ways to achieve this.  One might split the propagator in space-time 
into its expression within the forward light-cone and zero elsewhere, with the correction 
as the spacelike remainder.  Eq.(11c) shows another split, in momentum space, 
with the relativisitic  part equal to the usual propagator (recall that our definition 
(6) of $\phi^{-}$ is $(2M)^{1/2}$ times the usual definition) plus a part which is relatively small 
for $p^{0}\approx E$.
	 
	\section{Calculations}\label{Section IV} 

	The density matrix evolution equation and its perturbation series follow from Eq.(5b):
\begin{subequations}
\begin{eqnarray}\label{12}
\frac{d\rho(t)}{dt}&=&-(\lambda/2)\int d{\bf x}[A({\bf x},t),[A({\bf x},t),\rho(t)]]\\
\rho(t)&=&\sum_{n=0}^{\infty}\frac{(\lambda/2)^{n}}{n!}\int_{0}^{t} dx_{n}...\int_{0}^{t} dx_{1}
{\cal T}[A(x_{n}),...[A(x_{1}), \rho(t)]...].	
\end{eqnarray}
 \end{subequations}	
 \noindent  In this section, Eq.(12a) is used to calculate the rate for the wave function of a single free particle, in a superposition of two widely separated packets, to collapse to one of the packets.  Next, the 
 rate of energy increase for $N$ free particles is found. The following section discusses the formalism of QRCSL when particles are interacting.  Finally, this result is used to calculate, to first order in $\lambda$, the quadrupole 
 excitation rate of the $^{74}$Ge nucleus from its ground state to its first excited state. The result is compared with the present experimental upper limit on the rate of ``spontaneous" excitation in Ge.  
 
 \subsection{Collapse Rate for a Single Free Particle}
 
 	Consider a single particle initially in a superposition $|\psi, 0\rangle =\alpha|L\rangle +
 \beta |R\rangle$, where $|L\rangle$ and $|R\rangle$ are widely separated wavepackets, 
 so far apart that, to high accuracy, their regions of support do not overlap over the time interval $t$.  
 Let $|{\bf x}_{L}\rangle$ ($|{\bf x}_{R}\rangle$) be a position eigenstate 
 within the left (right) region of support.  The off-diagonal element of the density matrix, 
 using Eqs.(12a) and (7c), satisfies:
\begin{eqnarray}\label{13}
\frac{d\langle{\bf x}_{L}|\rho(t)|{\bf x}_{R}\rangle}{dt}&=&-(\lambda/2)\bigg( \frac{a^{2}}{4\pi^{3}}\bigg) ^{3/2}
M^{2}\int d{\bf x}\int\frac{d{\bf p}_{1}}{\sqrt{E_{1}}}\frac{d{\bf p}_{2}}{\sqrt{E_{2}}}\frac{d{\bf p}_{3}}{\sqrt{E_{3}}}\frac{d{\bf p}_{4}}{\sqrt{E_{4}}}
\nonumber\\
\qquad&\cdot&e^{-2^{-1}a^{2}[(p_{1}-(p_{2})^{2}+(p_{3}-(p_{4})^{2}]}
e^{-i(p_{1}-p_{2}+p_{3}-p_{4})\cdot x}\nonumber\\
\qquad&\cdot&\bigg\{ \delta ({\bf p}_{2}-{\bf p}_{3})\frac{1}{(2\pi)^{3}}\int d{\bf z}
[e^{i{\bf p}_{1}\cdot{\bf x}_{L}-i{\bf p}_{4}\cdot{\bf z}}\langle{\bf z}|\rho(t)| {\bf x}_{R}\rangle +
e^{i{\bf p}_{1}\cdot{\bf z}-i{\bf p}_{4}\cdot{\bf x}_{R}}\langle{\bf x}_{L}|\rho(t)| {\bf z}\rangle]\nonumber\\
\qquad&&-2\frac{1}{(2\pi)^{6}}\int d{\bf z}\int d{\bf z}'e^{i{\bf p}_{1}\cdot{\bf x}_{L}-i{\bf p}_{2}\cdot{\bf z}+
i{\bf p}_{3}\cdot{\bf z}'-i{\bf p}_{4}\cdot{\bf x}_{R}}\langle{\bf z}|\rho(t)| {\bf z}'\rangle
\bigg\}.
\end{eqnarray} 

	First, it can be seen that the last term in the curly brackets of Eq.(13) (arising from the 
term $-\lambda\int d{\bf x}A({\bf x},t)\rho(t)A({\bf x},t)$ in Eq. (12a)) may be neglected.  Consider the integral over ${\bf p}_{1}$,  ${\bf p}_{2}$ which appears in this term:
\begin{subequations}
\begin{eqnarray}\label{14}
f({\bf x}_{L}, {\bf x}, {\bf z};t)&\equiv&\int\frac{d{\bf p}_{1}}{\sqrt{E_{1}}}\frac{d{\bf p}_{2}}{\sqrt{E_{2}}}
e^{-2^{-1}a^{2}(p_{1}-p_{2})^{2}}e^{-i(p_{1}-p_{2}+p_{3}-p_{4})\cdot x}
e^{i{\bf p}_{1}\cdot{\bf x}_{L}-i{\bf p}_{2}\cdot{\bf z}}\\
&\approx&\sim e^{-(2a^{2})^{-1}({\bf x}-{\bf x}_{L})^{2}}
\delta ({\bf z}-{\bf x}_{L}),
\end{eqnarray}
 \end{subequations}	 
 \noindent where $\approx$ in (14b) means that we have set $t=0$ and $E_{1}= E_{2}=M$.  
 Even without the approximation, for any $t$,  if any one argument ({\bf x} or {\bf z}) of $f$ lies in $L$ and another in $R$, then 
 $f\approx 0$.  Now, the term under consideration has the form 
 \begin{subequations}
 \begin{eqnarray}\label{15}
&\sim& \int d {\bf x}d {\bf z}d {\bf z}' f({\bf x}_{L}, {\bf x}, {\bf z};t)f^{*}({\bf x}_{R}, {\bf x}, {\bf z}';t)
\langle{\bf z}|\rho(t)| {\bf z}'\rangle\\
&=&\int d {\bf x}d {\bf z}d {\bf z}' f({\bf x}_{L}, {\bf x}, {\bf z}_{L};t)f^{*}({\bf x}_{R}, {\bf x}, {\bf z}'_{R};t)
\langle{\bf z}_{L}|\rho(t)| {\bf z}'_{R}\rangle\approx 0.	
\end{eqnarray}
\end{subequations}	
\noindent In (15b),  ${\bf z}$ (${\bf z}'$) is restricted to ${\bf z}_{L}$ (${\bf z}'_{R}$) since, otherwise, $f$ ($f^{*}$) would vanish, and the result vanishes because either $f$ or $f^{*}$ vanishes  for 
every ${\bf x}$ .  

	In the remaining terms of (13), first perform the integral over ${\bf x}$, followed by the integrals over $p_{3}$, $p_{4}$, which results in 
\begin{eqnarray}\label{16}
\frac{d\langle{\bf x}_{L}|\rho(t)|{\bf x}_{R}\rangle}{dt}&=&-(\lambda/2)\bigg( \frac{a^{2}}{4\pi^{3}}\bigg) ^{3/2}
M^{2}\int \frac{d{\bf p}_{1}}{E_{1}}\frac{d{\bf p}_{2}}{E_{2}}
e^{-a^{2}(p_{1}-p_{2})^{2}}\nonumber\\
\qquad&&\int d{\bf z}
[e^{i{\bf p}_{1}\cdot({\bf x}_{L}-{\bf z})}\langle{\bf z}|\rho(t)| {\bf x}_{R}\rangle +
e^{-i{\bf p}_{1}\cdot({\bf x}_{R}-{\bf z})}\langle{\bf x}_{L}|\rho(t)| {\bf z}\rangle].
\end{eqnarray}
\noindent Next, the integral over ${\bf p}_{2}$ can be performed:
\begin{equation}\label{17}
\int \frac{d{\bf p}_{2}}{E_{2}}e^{-a^{2}(p_{1}-p_{2})^{2}}=(2\pi/a^{2})e^{2a^{2}M^{2}}K_{1}(2a^{2}M^{2}), 
\end{equation}
\noindent followed by the integral over ${\bf p}_{1}$:
\begin{equation}\label{18}
\int \frac{d{\bf p}_{1}}{E_{1}}e^{i{\bf p}_{1}\cdot({\bf x}_{L}-{\bf z})}=4\pi M
K_{1}(M|{\bf x}_{L}-{\bf z}|)/|{\bf x}_{L}-{\bf z}|. 
\end{equation}

	The evolution equation now reads:
\begin{eqnarray}\label{19}
\frac{d\langle{\bf x}_{L}|\rho(t)|{\bf x}_{R}\rangle}{dt}&=&-(\lambda a M^{3}/2\pi^{5/2})
e^{2a^{2}M^{2}}K_{1}(2a^{2}M^{2})\nonumber\\
&\cdot&\int d{\bf z}\big[ \frac{K_{1}(M|{\bf x}_{L}-{\bf z}|)}{|{\bf x}_{L}-{\bf z}|}
\langle{\bf z}|\rho(t)| {\bf x}_{R}\rangle + \frac{K_{1}(M|{\bf x}_{R}-{\bf z}|)}{|{\bf x}_{R}-{\bf z}|} 
\langle{\bf x}_{L}|\rho(t)| {\bf z}\rangle\big].
\end{eqnarray} 
\noindent Eq. (19) is exact.  

	Now, specialize to the case where $a$ is much larger than 
the particle Compton wavelength $M^{-1}$ so 
\begin{equation}\label{20}
e^{2a^{2}M^{2}}K_{1}(2a^{2}M^{2})\approx(2aM)^{-1}\pi^{1/2}.
\end{equation}
\noindent Further, if the wave packets change slowly  on the distance scale of $M^{-1}$ then,  in 
the integrand of Eq.(19),
$K_{1}(M|{\bf x}-{\bf z}|)/ |{\bf x}-{\bf z}|$ may be well approximated by a delta-function, whose numerical coefficient may be found by integrating Eq.(18) over ${\bf z}$:
\begin{equation}\label{21}
K_{1}(M|{\bf x}-{\bf z}|)/|{\bf x}-{\bf z}|\approx (2\pi^{2}/M^{2})\delta ({\bf x}-{\bf z}) 
\end{equation}

	The result of inserting (20), (21) into (19) gives the result:  
	\begin{equation}\label{22}
\frac{d\langle{\bf x}_{L}|\rho(t)|{\bf x}_{R}\rangle}{dt}=-\lambda \langle{\bf x}_{L}|\rho(t)|{\bf x}_{R}\rangle,
\end{equation}
\noindent i.e., rate of decay of the off-diagonal density matrix element is $\lambda$, 
the same rate as for nonrelativistic CSL.

\subsection{Energy Creation Rate For Free Particles}	
	Calculation of the energy creation rate for n free particles begins by multiplying 
Eq.(12a) by the free Hamiltonian $H_{0}$ and taking the trace:
\begin{subequations}
 \begin{eqnarray}\label{23}
 \frac{d{\bar H}}{dt}&=&-\frac{\lambda}{2}\int d{\bf x}\hbox {Tr}\{ 
 [A(x),[A(x),H_{0}]]\rho(t)\}\\
 &=&-\frac{\lambda i}{2}\int d{\bf x}\hbox {Tr}\{ 
 [A(x),{\dot A}(x)]\rho(t)\}, 
 \end{eqnarray}
\end{subequations}
\noindent where ${\bar H}(t)\equiv\hbox{Tr}\{ H_{0}\rho(t)\}$.  Using the expression (7c) for $A(x)$, the commutator can be evaluated.  Then, 
the integral over ${\bf x}$ can be performed and, using the resulting delta-functions 
of momentum differences, two of the momentum integrals can be performed, resulting in 
\begin{subequations}
 \begin{eqnarray}\label{24}
 &&\frac{d{\bar H}(t)}{dt}=\frac{\lambda M^{2}a^{3}}{2\pi^{3/2}}
 \int\frac{d{\bf p}_{1}}{E_{1}}\frac{d{\bf p}_{2}}{E_{2}}(E_{2}-E_{1})e^{-a^{2}(p_{1}-p_{2})^{2}}\nonumber\\
&&\qquad\qquad\qquad\qquad \qquad\cdot\hbox{Tr}\{[a^{\dagger}({\bf p}_{1})a({\bf p}_{1})-a^{\dagger}({\bf p}_{2})a({\bf p}_{2})]\rho (t)\}\\
&&\negthinspace\negthinspace\negthinspace\negthinspace\negthinspace\negthinspace
=\frac{\lambda M^{2}a^{3}}{\pi^{3/2}}\hbox{Tr}\{ [ \int\frac{d{\bf p}_{1}}{E_{1}}a^{\dagger}({\bf p}_{1})a({\bf p}_{1})\int d{\bf p}_{2}e^{-a^{2}(p_{1}-p_{2})^{2}}-N\int\frac{d{\bf p}_{2}}{E_{2}}e^{-a^{2}(p_{1}-p_{2})^{2}}]\rho (t) \}
\end{eqnarray}
\end{subequations}
\noindent  In Eq.(24b), 
$N\equiv\int d{\bf p}_{1}a^{\dagger}({\bf p}_{1})a({\bf p}_{1})$ is the number-of-particles operator.

	The last integral in Eq.(24b) over ${\bf p}_{2}$ is given in Eq.(17).  The first  integral 
over ${\bf p}_{2}$ is
\begin{equation}\label{25}
\int d{\bf p}_{2}e^{-a^{2}(p_{1}-p_{2})^{2}}=(2\pi/a^{2})E_{1}e^{2a^{2}M^{2}}
[K_{0}(2a^{2}M^{2})+(a^{2}M^{2})^{-1}K_{1}(2a^{2}M^{2})]
\end{equation}
\noindent Because (25) is proportional to $E_{1}$, the first integral in Eq.(24b) over ${\bf p}_{1}$ is 
$\sim N$.  Since $\hbox{Tr}\{N\rho(t)\}=n\hbox{Tr}\rho(t)=n$, one obtains the exact result that 
 ${\bar H}(t)$ increases linearly with time: 
 \begin{equation}\label{26}
\frac{d{\bar H}(t)}{dt}=\lambda n M^{2}a2\pi^{-1/2}e^{2a^{2}M^{2}}
[K_{0}(2a^{2}M^{2})-K_{1}(2a^{2}M^{2})(1-(a^{2}M^{2})^{-1})].
\end{equation}

	In the case where $aM>>1$, using the series expansions
\begin{equation}\label{27}
K_{0}(z)=\sqrt{\frac{\pi}{2z}}e^{-z}\bigg[1-\frac{1}{8z}\bigg],\qquad 
K_{1}(z)=\sqrt{\frac{\pi}{2z}}e^{-z}\bigg[1+\frac{3}{8z}\bigg],
\end{equation}
\noindent Eq.(26) becomes:
\begin{equation}\label{28}
\frac{d{\bar H}(t)}{dt}=\frac{3\lambda n}{4 M a^{2}},
\end{equation}
\noindent the same energy creation rate as for nonrelativistic CSL. 

\section{QRCSL for Interacting Particles}
	
	The derivative of Eq.(1),
\begin{equation}\label{29}
d|\psi , t \rangle_{w}/dt= -(4\lambda)^{-1}\int d{\bf x}[w(x)-2\lambda A( x)]^{2}
|\psi , t \rangle_{w}
\end{equation}
\noindent where $A(x)$ is given by Eqs.(7),  is the QRCSL evolution equation for 
noninteracting particles in the ``collapse interaction picture," where the field 
operators evolve freely and the state vector changes only due to the collapse evolution. 
As usual, the Schr\"dinger picture statevector is
$|\psi , t \rangle_{w}^{s}=\exp-iH_{0}t|\psi , t \rangle_{w}$, and its evolution equation 
follows from Eq.(29):
\begin{equation}\label{30}
d|\psi , t \rangle_{w}^{s}/dt= -iH_{0}|\psi , t \rangle_{w}^{s}-(4\lambda)^{-1}\int d{\bf x}
[w(x)-2\lambda A({\bf x},0)]^{2}|\psi , t \rangle_{w}^{s},
\end{equation}
\noindent where $A({\bf x},0)$ is given by Eqs.(7) with $t=0$ (so, in particular, 
the operators in (7b) still have $ib_{0}$ as time arguments).  In the Schr\"odinger picture, 
the operators do not evolve, and the state vector changes with time due to the free evolution as well as 
due to the collapse evolution.
	  
	  	As usual, to add interaction, one replaces $H_{0}$ by $H=H_{0}+V$ in Eq.(30). Transforming back to the interaction picture gives:
\begin{equation}\label{31}
d|\psi , t \rangle_{w}/dt= -iV(t)|\psi , t \rangle_{w}-(4\lambda)^{-1}\int d{\bf x}
[w(x)-2\lambda A( x)]^{2}|\psi , t \rangle_{w},
\end{equation}
\noindent where $V(t)\equiv \exp (iH_{0}t)V\exp -(iH_{0}t)$ is a Lorentz scalar, the four-integral of a local scalar density.  However, although 
Eq.(31) is form covariant, it is not Lorentz invariant because, not only doesn't $A$ commute with 
itself at space-like separation, it usually will not commute with the local scalar density at space-like separation either.  
However, such a local density is constructed from $\phi=\phi^{+}+\phi^{-}$, so its commutator with 
$A$ falls off exponentially with space-like separation as in (10).  
In this sense this interacting QRCSL model is also ``quasi-relativistic."  In this interaction picture, 
the quantum fields evolve freely and the state vector evolves due to the interaction and the collapse.   

	For some calculations, it is useful to work in the collapse-interaction 
picture, where the fields are Heisenberg fields, evolving according to 
the interacting quantum field theory and the state-vector only changes with time due to 
the collapse evolution.  Using (31) to go to the Schr\"odinger picture and, as usual,  defining  the collapse-interaction picture statevector as $|\psi , t \rangle'_{w}=\exp (iHt)|\psi , t \rangle_{w}^{s}$, one obtains 
\begin{equation}\label{32}
d|\psi , t \rangle'_{w}/dt= -(4\lambda)^{-1}\int d{\bf x}[w(x)-2\lambda A'( x)]^{2}
|\psi , t \rangle'_{w},
\end{equation}
\noindent where $A'( x)=\exp( iHt)A({\bf x},0)\exp-(iHt)$.  It is worth emphasizing that 
$A'( x)$ is {\it not} given by Eq.(7b) with operators 
$\phi^{\pm}({\bf x}+{\bf b}, t + ib_{0})$ 
replaced by $\exp [iH(t+ib_{0}) ]\phi^{\pm}({\bf x}+{\bf b},0) \exp -[iH(t+ib_{0}) ]$ 
but, rather, is composed of operators $\exp (iHt) \phi^{\pm}({\bf x}+{\bf b},ib_{0}) \exp -(iHt)$, according 
to this prescription. 

\subsection{Bound State Excitation to Lowest Order}
	The density matrix evolution equation which follows from Eq.(32) is 
Eq.(5b) with $A$ replaced by $A'$. To lowest order in $\lambda$, 
this is
\begin{equation}\label{33}
\rho(t)=\rho(0) -(\lambda/2)\int_{0}^{t} dtd{\bf x}[A'^{2}({\bf x},t)\rho(0) 
+\rho(0)A'^{2}({\bf x},t)-2A'({\bf x},t))\rho(0) A'({\bf x},t)].
\end{equation}
\noindent Time ordering plays no role to lowest order, so the 
expression (33) is Lorentz invariant.  Take $\rho (0)=|E_{i}\rangle \langle E_{i}|$, where 
$|E_{i}\rangle$ is a bound $N$-particle state that is an energy eigenstate with eigenvalue $E_{i}$, and is also an eigenstate with eigenvalue 0 of the center of mass operator 
${\bf Q}\equiv \sum_{n=1}^{N}{\bf X}_{n}/N$ (${\bf X}_{n}$ is the position operator of the $n$th 
particle). It is desired to calculate the probability that the system is excited to 
the bound energy eigenstate $|E_{f}\rangle$ ($H|E_{f}\rangle=E_{f}|E_{f}\rangle$
and ${\bf Q}|E_{f}\rangle=0$).  Because $\langle E_{i}|E_{f}\rangle=0$, only the last term in 
(33) contributes:  
\begin{equation}\label{34}
\langle E_{f}|\rho(t)|E_{f}\rangle=\lambda\int_{0}^{t} dtd{\bf x}
\langle E_{f}|A'({\bf x},t)|E_{i}\rangle \langle E_{i}| A'({\bf x},t)|E_{f}\rangle.
\end{equation}
\noindent Moreover, since $\langle E_{f}|A'({\bf x},t)|E_{i}\rangle=\exp i( E_{f}-E_{i})
\langle E_{f}|A({\bf x},0)|E_{i}\rangle$, the integrand 
in (34) is time independent, so the excitation rate $\Gamma\equiv d\langle E_{f}|\rho(t)|E_{f}\rangle/dt$ is constant:
\begin{equation}\label{35}
\Gamma=\lambda\int d{\bf x}
|\langle E_{f}|A({\bf x},0)|E_{i}\rangle |^{2}.
\end{equation}

	At this point,  assume that the particles in the initial and final states 
move nonrelativistically, $({\bf p}/mc)^{2}<<1$, so that, in the expression (7c) 
for $A(x)$, one can make the approximations $E\approx E'\approx M$ 
and $(p-p')^{2}\approx ({\bf p}-{\bf p}')^{2}$. Then $A({\bf x},0)$ in Eq.(35) 
becomes the non-relativistic expression (2). For 
completeness,  the analysis leading to Eq.(37) shall be given here, instead of just quoted\cite{PearleSquires, Collett}.  Start by noting that 
$\int d{\bf x}'F({\bf x}')\xi^{\dagger}({\bf x}')\xi({\bf x}')=\sum_{n=1}^{N}F({\bf X}_{n})$ where 
$F$ is an arbitrary function.  Use of the form (2b) for $A({\bf x},0)$ in Eq.(35) results in:
\begin{subequations}
\begin{eqnarray}\label{36}
\Gamma &=&\lambda (\pi a^{2})^{-3/2}\int d{\bf x}
|\langle E_{f}|\int d{\bf x}' e^{-(2a^{2})^{-1}({\bf x}-{\bf x}')^{2}}\xi^{\dagger}({\bf x}')
\xi({\bf x}')|E_{i}\rangle |^{2}\\
&=&\lambda (\pi a^{2})^{-3/2}\int d{\bf x}|\langle E_{f}|\sum_{n=1}^{N} e^{-(2a^{2})^{-1}({\bf x}-{\bf X}_{n})^{2}}|E_{i}\rangle |^{2}\\
&=&\lambda\sum_{n,m=1}^{N}\langle E_{f}|\big\{e^{-(4a^{2})^{-1}({\bf X}_{nL}-
{\bf X}_{mR})^{2}}|E_{i}\rangle \langle E_{i}|\big\} | E_{f}\rangle
\end{eqnarray}
\end{subequations}

	Next, expand the expression in (36c) in a power series in (size of bound state/$a)^{2}$.  
Then, because of the orthogonality of the initial and final states and because ${\bf Q}|E_{i,f}\rangle=0$,
the first nonvanishing term of (36c) is of order $a^{-4}$§:
\begin{equation}\label{37}
\Gamma=\lambda(2a)^{-4}\bigg [|\langle E_{f}|\sum_{n=1}^{N}{\bf X}_{n}^{2}|E_{i}\rangle|^{2}+
2\sum_{n=1,m=1}^{N}\sum_{i,j=1}^{3}\langle E_{f}|{\bf X}_{n}^{i}{\bf X}_{n}^{j}|E_{i}\rangle
\langle E_{i}|{\bf X}_{m}^{i}{\bf X}_{m}^{j}|E_{f}|\rangle\bigg ].
\end{equation}

\subsection{Excitation of $^{74}$Ge Nucleus}

	Now, Eq.(37) is to be applied to collapse-induced spontaneous excitation of a proton from the ground state ($0^{+})$ of 
a $^{74}$Ge nucleus (the largest percentage isotope-36.5\%-in naturally occurring Ge) to 
its first excited state ($2^{+}$) at .596MeV\cite{Nuc}.  The experimental upper limit on 
spontaneous emission of  .596MeV gammas in Ge, obtained by  observing the radiation 
from an isolated slab of Ge for a long time, is $\approx .03$ counts/kg-day (in a 2 MeV bin)
\cite{Avignone}.  
The analysis for spinless particles of one mass given above can be applied to this case because the proton and neutron have almost the same mass and there is no spin-flip involved in this transition.   

	One readily finds from (37) the expression for the quadrupole excitation rate:
\begin{equation}\label{38}
\Gamma=(\pi/15)(\lambda/a^{4})\sum_{m,m'=-2}^{2}|\langle 2^{+},m'|\sum_{n=1}^{Z}{\bf X}_{n}^{2}Y_{2m}(\Theta_{n}, \Phi_{n})|0^{+}\rangle|^{2},  
\end{equation}
\noindent where $\Theta_{n}$, $\Phi_{n}$ are angle operators for the $n$th particle and $Y_{2m}$
is a spherical harmonic.  

	Now, the  lifetime $\tau$ of the $2^{+}$ state is given by the expression\cite{BW}
\begin{equation}\label{39}
\tau^{-1}=(4\pi/3\cdot 5^{3})ck^{5}(e^{2}/\hbar c)\sum_{m,m'=-2}^{2}|\langle 2^{+},m'|\sum_{n=1}^{Z}{\bf X}_{n}^{2}Y_{2m}(\Theta_{n}, \Phi_{n})|0^{+}\rangle|^{2} ,  
\end{equation}
\noindent where $k\approx 3.2\cdot10^{10}$cm$^{-1}$ is the .596 MeV photon wavenumber.  From 
Eqs. (38), (39) is obtained 
\begin{equation}\label{40}
\Gamma_{QRCSL}=(5/2)^{2}\lambda [(ak)^{4}(e^{2}/\hbar c)kc\tau]^{-1}\approx 5\cdot 10^{-16}
\hbox{counts/kg-day},
\end{equation}
\noindent with use of the numbers $\tau=17.9$psec as the experimental lifetime of the state, $\approx 8.3\cdot 10^{24}$ as the number of nucleii/kg of Ge (so there are $\approx 3.0\cdot 10^{24}$ 
$^{74}$Ge nucleii/kg), and $8.6\cdot 10^{4}$sec/day.  $\Gamma_{QRCSL}$ is well below the experimental 
upper limit of $3\cdot 10^{-2}$counts/kg-day.

	This contrasts with the situation for the relativistic collapse model RCSL.  From Eqs.(A4) and (39)  
is obtained
\begin{equation}\label{41}
\Gamma_{RCSL}=(5/3\pi^{2})\lambda  a[(e^{2}/\hbar c)c\tau]^{-1}\approx 5\cdot 10^{10}
\hbox{counts/kg-day}, 
\end{equation}
\noindent which far exceeds the experimental upper limit.  These calculations were performed assuming the GRW values for $\lambda$,  $a$ but, for most of the range of these parameters allowed by other considerations\cite{Collett}, $\Gamma_{RCSL}$ is still excessive.  The reason for the difference in excitation 
between QRCSL and RCSL is that, in the former, just as in nonrelativistic CSL, collapse narrows the excited particle's wavefunction to $a$ whereas, in the latter, collapse narrows the wavefunction 
to $[a^{-2}+(E_{f}-E_{i})^{2}]^{-1/2}$.

\section{Concluding remarks}\label{Section V}

	Because all  previous CSL-type relativistic collapse models except RCSL are untenable since they produce infinite energy/sec-vol from the vacuum, and RCSL produces too much nuclear excitation, the QRCSL model has been suggested. It has form-invariant equations, but it fails to be relativistic because its Lorentz invariant operators do not commute at space-time separation.  However, since these operators ``almost" commute, I believe that such quasi-relativistic behavior is worth consideration, as a close and experimentally testable variant of special relativity combined with a description of collapse that is as close as could be expected to nonrelativistic CSL.  

	However, the model is, after all, described in a preferred frame,  the one where the time-ordering operation is defined.  One might tentatively identify the preferred frame with the local co-moving frame of the universe\cite{Stapp}.  Exploration of the extent of violation of Lorentz invariance for various hypothetical situations is certainly of interest.  Since QRCSL's slow speed limit is CSL which, so far, has defied experimental refutation, it may be worthwhile to examine schemes whereby detectors move at high speeds.  One may also examine whether the frame dependent, although 
non-detectable, wave packet collapse 
locales and times (e.g.,  in EPR-type situations) in relativistic collapse models are similar to those of QRCSL, or if  the preferred 
frame's wave packet collapse locales and times might, in some sense,{\it be} preferred

\appendix
\section{Bound State Excitation in RCSL}\label{App A}
	In RCSL, the only finite relativistic collapse model extant, the expression for the excitation probability comparable to Eq.(34) (but here taken in the nonrelativistic limit)  is 
\begin{equation}\label{A1}
\langle E_{f}|\rho(t)|E_{f}\rangle=4\lambda a\int_{0}^{t} dxdx'G(x-x')
\langle E_{f}|\xi^{\dagger}(x)\xi(x)|E_{i}\rangle \langle E_{i}| \xi^{\dagger}(x')\xi(x')|E_{f}\rangle.
\end{equation}
\noindent In Eq.(A1), $G(x-x')=(2\pi)^{-4}\int dp\exp ip\cdot (x-x')\delta (p^{2}-a^{-2})$, i.e., this is a 
non-Markovian model with a tachyonic noise spectrum whose ``tachyon mass" is $a^{-1}\approx 2$eV.  
The operator $\xi^{\dagger}(x)=\exp iHt \xi^{\dagger}({\bf x},0)\exp - iHt$, with $\xi^{\dagger}({\bf x},0)$ 
given by Eq.(3), is the Heisenberg creation operator.  
Use this first, to pull out the time dependence from the matrix elements in (A1), letting  $H$ act on the energy eigenstates, and then perform the time integrals with the result
\begin{eqnarray}\label{A2}
&&\negthinspace\negthinspace\negthinspace\negthinspace\negthinspace\negthinspace
\negthinspace\negthinspace\negthinspace\negthinspace\negthinspace\negthinspace
\langle E_{f}|\rho(t)|E_{f}\rangle=4(2\pi)^{-4}\lambda a\int_{0}^{t} dxdx' dp
 e^{i{\bf p}\cdot ({\bf x}-{\bf x}')}
\delta (p^{2}-a^{-2})\nonumber\\
&&\negthinspace\negthinspace\negthinspace\negthinspace\negthinspace\negthinspace
\cdot\bigg\{ \sin[(E_{f}-E_{i}-p^{0})t/2]/[(E_{f}-E_{i}-p^{0})/2\big]\bigg\}^{2}
 \langle E_{f}|\xi^{\dagger}({\bf x})\xi({\bf x})|E_{i}\rangle \langle E_{i}| \xi^{\dagger}
 ({\bf x}')\xi({\bf x}')|E_{f}\rangle.
\end{eqnarray}
\noindent For large $t$, $[\sin(\alpha t)/\alpha]^{2}\approx t\pi \delta(\alpha)$. Then, using this delta function to perform the 
integral over $p^{0}$, and utilizing $\int d{\bf x}'F({\bf x}')\xi^{\dagger}({\bf x}')\xi({\bf x}')=\sum_{n=1}^{N}F({\bf X}_{n})$ as was done in obtaining Eq.(36), we obtain
\begin{subequations}
\begin{eqnarray}\label{A3}
&&\negthinspace\negthinspace\negthinspace\negthinspace\negthinspace\negthinspace
\negthinspace\negthinspace\negthinspace\negthinspace\negthinspace\negthinspace
\negthinspace\negthinspace\negthinspace\negthinspace\negthinspace\negthinspace
\negthinspace\negthinspace\negthinspace\negthinspace\negthinspace\negthinspace
\Gamma=\lambda a2^{-1}\pi^{-3}\int d{\bf p}\delta [{\bf p}^{2}-(E_{f}-E_{i})^{2}-a^{-2}]\sum_{n,m=1}^{N}
\langle E_{f} |\big\{ e^{i{\bf p}\cdot ({\bf X}_{nL}-{\bf X}_{mR})
} |E_{i} \rangle \langle E_{i}| \big\} |E_{f}\rangle\\
&=&\lambda a\pi^{-2}\sum_{n,m=1}^{N}
\langle E_{f} |\big\{ [\sin(k|{\bf X}_{nL}-{\bf X}_{mR}|)/|{\bf X}_{nL}-{\bf X}_{mR}|]
 |E_{i} \rangle \langle E_{i}| \big\} |E_{f}\rangle.
\end{eqnarray}
\end{subequations}
In Eq.(A3b), $k=\sqrt{(E_{f}-E_{i})^{2}+a^{-2}}\approx E_{f}-E_{i}$ (if  $E_{f}-E_{i} >>a^{-1}$) is 
the wavenumber of a photon making the transition from the excited state to the ground state.  

	Eq.(A3b) may be compared to Eq. (36c).  The gaussian with width $a$ there, 
is replaced by the $\sin kz/z$ form with width $k^{-1}$ here.  An expansion in powers of 
$a^{-1}$ there is replaced by an expansion in powers of $k$ here.  It is because 
$k>>a^{-1}$ that this RCSL model produces a much larger excitation rate than 
QRCSL.  The first non-vanishing term in Eq.(A3b) is 
\begin{equation}\label{A4}
\Gamma=2\lambda a k^{5}(5! \pi^{2})^{-1}\bigg [|\langle E_{f}|\sum_{n=1}^{N}{\bf X}_{n}^{2}|E_{i}\rangle|^{2}+
2\sum_{n=1,m=1}^{N}\sum_{i,j=1}^{3}\langle E_{f}|{\bf X}_{n}^{i}{\bf X}_{n}^{j}|E_{i}\rangle
\langle E_{i}|{\bf X}_{m}^{i}{\bf X}_{m}^{j}|E_{f}|\rangle\bigg ].
\end{equation}
\noindent This is identical to Eq.(37) except for the numerical factor, and is 
used in Eq.(41).


\begin{thebibliography}{99}
\bibitem{PearleCSL} P. Pearle, Phys. Rev. A {\bf 39}, 2277 (1989). 
\bibitem{GPR} G. C. Ghirardi, P. Pearle and A. Rimini, Phys. Rev. A {\bf 42}, 78 (1990). 
\bibitem{PearleRel} P. Pearle, in {\it Sixty-Two Years of Uncertainty}, edited by A. Miller, 
(Plenum, New  York 1990), p. 193.
\bibitem{GGP}	G. C. Ghirardi, R. Grassi and P. Pearle, Found. Phys. {\bf 20}, 1271 (1990);
\bibitem{PearleRel2} P. Pearle in {\it Quantum Chaos-Quantum Measurement}, 
edited by P. Cvitanovic et. al., (Kluwer, the Netherlands 1992), p.283.
\bibitem{Pearleenergy}P. Pearle, Found. Phys. {\bf 30}, 1145 (2000).
\bibitem{experiment} B. Collett, P. Pearle, F. Avignone and S. Nussinov, 
 Found. Phys. {\bf 25}, 1399 (1995); P. Pearle, J. Ring, J. I. Collar and F. Avignone,
Found. Phys. {\bf 29}, 465 (1999); G. Jones, P. Pearle and J. Ring, Found. Phys. {\bf 34}, 1467 (2004).
\bibitem{PearleNoise} P. Pearle, Phys. Rev. A {\bf 48}, 913 (1993); in {\it Stochastic Evolution of Quantum States in Open Systems and in Measurement Processes}, edited by L. Diosi 
and B. Lukacs (World Scientific, Singapore 1994), p. 79; 
in {\it Perspectives on Quantum Reality}, edited by R. Clifton (Kluwer, Dordrecht 1996), p. 93.       
\bibitem{BGNoise}A. Bassi and G.C. Ghirardi, Phys. Rev. A{\bf 65}, 042114 (2002).
\bibitem{PearleTachyon}P. Pearle, Phys. Rev. A{\bf 59}, 80 (1999).
\bibitem{RN} O. Nicrosini and A. Rimini, Found. Phys. {\bf 33}, 1061 (2003).  This manifestly covariant CSL-type model is quasi-relativisitic in that, like QRCSL, the collapse-inducing operator does not commute with 
itself at spacelike separation.  However, it is not quasi-relativistic since, unlike QRCSL, this commutator does not decrease rapidly with increasing distance.  This model violates 1), 2), 4) and 5).   
Although nonrelativistic and Markovian  with a nonlocal collapse-inducing operator, with a slight alteration it can be cast as relativistic  and non-Markovian with a local collapse-inducing operator. In this form it violates 3), 4) and 5). 
\bibitem{Tumulka} C. Dove and E. J. Squires, {\it A Local Model of Explicit Wavefunction Collapse},  quant-ph /0406094, was the first attempt at a relativistic generalization of GRW's "hitting" process.  R. Tumulka, {\it A Relativistic Version of the Ghirardi-Rimini-Weber Model}, quant-ph /0406094 has recently presented an ingenious improved model, for free particles satisfying the Dirac equation. It converts free particles to antiparticles, so it violates 5).  
\bibitem{GRW} G. C. Ghirardi, A. Rimini and T. Weber, Phys. Rev. D {\bf 34}, 470 (1986); 
Phys. Rev. D {\bf 36}, 3287 (1987); 
Found. Phys. {\bf 18}, 1, (1988). 
\bibitem{Collett} B. Collett, P. Pearle, F. Avignone and S. Nussinov in reference \cite{experiment};
 B. Collett and P. Pearle, Found. Phys. {\bf 33}, 1495 (2003). 
\bibitem{Namsrai}A. Pais and G. E. Uhlenbeck, Phys. Rev. {\bf 79}, 145 (1950); G. V. Efimov, JETP {\bf 44}, 2107 (1963); for a comprehensive discussion and citation of other papers, see K. Namsrai,  
 {\it Nonlocal Quantum Field Theory and Stochastic Quantum Mechanics}, (Reidel, Dordrecht, 1986).
 \bibitem{PearleSquires}	P. Pearle and E. Squires, Phys. Rev. Lett. {\bf 73}, 1 (1994).
 \bibitem{Nuc} National Nuclear Data Center, www.nndc.bnl.gov/index.jsp.
 \bibitem{Avignone}H. S. Miley, F. T. Avignone III, R. L. Brodzinski, J. I. Collar and 
J. H. Reeves, Phys. Rev. Lett. {\bf 65}, 3092 (1990).
 \bibitem{BW}J. M. Blatt and V. F. Weisskopf, {\it Theoretical Nuclear Physics} (Wiley, New York 1952), p.595.
 \bibitem{Stapp}I first heard this suggestion a decade ago from Henry Stapp (private communication),
 and was amazed that such a distinguished  relativisitic quantum field theorist would not be disconcerted  
 if it proved impossible to make a viable relativistic collapse model. 
 
\end{thebibliography}
\end{document}